\documentclass[letterpaper, 10 pt, conference]{ieeeconf}
\IEEEoverridecommandlockouts
\usepackage{cite}
\usepackage{amsmath,amssymb,amsfonts,ulem,lipsum}
\usepackage{algorithmic}
\usepackage{graphicx,svg}
\usepackage{textcomp}
\usepackage{xcolor}
\usepackage{framed}
\usepackage{mdframed}
\def\BibTeX{{\rm B\kern-.05em{\sc i\kern-.025em b}\kern-.08em
    T\kern-.1667em\lower.7ex\hbox{E}\kern-.125emX}}

\newcommand{\calA}{{\mathcal{A}}}
\newcommand{\calF}{{\mathcal{F}}}

\newcommand{\calL}{{\mathcal{L}}}

\newcommand{\KK}{{\mathbb{K}}}

\newcommand{\RR}{{\mathbb{R}}}
\newcommand{\NN}{{\mathbb{N}}}

\newcommand{\XX}{{\mathbb{X}}}

\newcommand{\knl}{\mathfrak{K}}

\newcommand{\unl}{\ell}
\newcommand{\wnl}{w}

\newcommand{\smallT}{\scriptscriptstyle\mathsf{T}}

\newcommand{\Pwr}{\mathcal{P}}

\let\oldref\ref
\renewcommand{\ref}[1]{(\oldref{#1})}

\newtheorem{theorem}{Theorem}
\newtheorem{corollary}{Corollary}

\title{Rates of Convergence  in  a Class of Native Spaces\\ for Reinforcement  Learning and Control}

\author{Ali Bouland$^{1}$, Shengyuan Niu$^{1}$, Sai Tej Paruchuri$^{2}$, Andrew Kurdila$^{1}$, John Burns$^{3}$, Eugenio Schuster$^{2}$ 
\thanks{$^{1}$Ali Bouland, Shengyuan Niu and Andrew Kurdila are with the Department of Mechanical Engineering, Virginia Tech, Blacksburg, VA 24060, USA {\tt\small bouland@vt.edu}}%
\thanks{$^{2}$Sai Tej Paruchuri and Eugenio Schuster are with the Department of Mechanical Engineering and Mechanics, Lehigh University, Bethlehem, PA 18015, USA {\tt\small saitejp@lehigh.edu}
}
\thanks{$^{3}$John Burns is with the Department of Mathematics, Virginia Tech, Blacksburg, VA 24060, USA {\tt\small jaburns@vt.edu} }%
}

\begin{document}

\maketitle

\begin{abstract}
This paper studies convergence rates for  some  value function approximations that arise in a collection of  reproducing kernel Hilbert spaces (RKHS) $H(\Omega)$. By casting an optimal control problem in a specific class of native spaces, strong rates of convergence are derived for the operator equation that enables offline approximations that appear in policy iteration. Explicit upper bounds on error in value function and controller approximations are derived in terms of power function $\Pwr_{H,N}$ for the space of finite dimensional approximants $H_N$  in the native space $H(\Omega)$. These bounds are geometric in nature and refine some well-known, now classical results concerning convergence of approximations of value functions. 
\end{abstract}

\section{Introduction}
\label{ssec_probstmt}
Consider a nonlinear system that is governed by the ordinary differential equations 
\begin{align}
\label{eq_sys}
    \dot{x}(t) = f(x(t)) + g(x(t)) u(t), \qquad x(0) = x_0, 
\end{align}
where $x(t) \in \RR^n$ is the state, $u(t) \in \RR^m$ is the input,  and $f : \RR^n \to \RR^n$, $g : \RR^n \to \RR^{n \times m}$ are known functions. We assume $f(0)=0$ and are interested in a regulation problem that drives this system to the origin. 


We seek an admissible state feedback function $\mu : x \mapsto \mu(x)$ that can stabilize the system presented above. Often, we restrict consideration to feedback functions $\mu$ that leave some subset of interest $\Omega$ positive invariant. In addition to stabilizing the system, a feedback function  must then be continuous on  $\Omega$ and satisfy $\mu(0) = 0$ to consider it admissible on the subset $\Omega$ of the state space. The cost associated with an admissible control policy $\mu$ is consequently defined as
\begin{align}
    V_\mu(x_0) = \int_0^\infty r(x(\tau),\mu(x(\tau)) d \tau,
\end{align}
where $r(x,\mu) = Q(x) + \mu^\mathsf{T} R \mu$, $Q(x)$ is a positive definite function and $R$ is a symmetric positive definite matrix.
Assuming that the function $V_\mu(x): x \to V_\mu(x)$ is continuously differentiable, we can write down its differential Lyapunov-like equation in terms of the Hamiltonian as 
\begin{align}
    \mathcal{H}(x, &\mu, \nabla V)\label{eq:hamiltonian}\\
    &:=\underbrace{\left( f(x) + g(x) \mu(x) \right)^\mathsf{T} \nabla V_\mu(x)}_{:=(AV_\mu)(x)} + r(x,\mu(x)) = 0 \nonumber
\end{align}
where $\nabla$ denotes the gradient operator. The goal of optimal control is to choose a control policy $\mu^*$ such that $V_{\mu^*}(x_0)$ is minimized.
The function  $V_{\mu^*}$ is commonly referred to as the value function.  Standard optimal control analysis \cite{bertsekas2012dynamic,bertsekas1997nonlinear,lewis2013reinforcement} shows that the value function satisfies the Hamilton-Jacobi-Bellman (HJB) equation
\begin{equation}
    0 = min_{\mu \in M(\Omega)} \mathcal{H}(x, \mu, \nabla V_{\mu^*}),
\end{equation}
which is equivalent to
$
    0 = \mathcal{H}(x^*, \mu^*, \nabla V_{\mu^*}), 
$
where $\mu^*$ is given by
$
    \mu^* = - \frac{1}{2} R^{-1} g^{\smallT} \nabla V_{\mu^*}, 
$
and $x^*$ is the optimal trajectory generated by $\mu^*$. Once the HJB equation is solved for the optimal value function, the optimal controller can be found using this equation for $\mu^*$. 

In general, the HJB equation is a nonlinear partial differential equation that is difficult to solve, and the technical literature that studies this problem is vast. 
 Among this collection of work, a few ``now-classic'' papers related to the study of Galerkin approximations are particularly relevant to this paper. These include the notable early efforts in \cite{bea1998successive,beard1997galerkin}. The highly cited work in \cite{abu2005nearly} builds on the earlier work on Galerkin approximations to handle saturating actuators, which is subsequently used to form the theoretical foundation in \cite{vamvoudakis2010online}  and many subsequent works \cite{vamvoudakis2021handbook,kiumarsi2017optimal,lewis2013reinforcement,kamalapurkar2018reinforcement}.

The treatises \cite{lewis2013reinforcement} and \cite{kamalapurkar2018reinforcement} give excellent accounts of the theory for reinforcement learning (RL) methods, and recent surveys include \cite{vamvoudakis2021handbook,kiumarsi2017optimal}. One popular method of approximating the solution of the HJB equation is the actor-critic method. It entails an iterative approach of approximating the value function using the critic, then the actor uses the value approximation to get a control policy estimate, and the process repeats.  A second common method is policy iteration (PI), which requires full knowledge of the system dynamics but allows an offline calculation of the optimal control law. The effectiveness of both  methods relies  on the convergence of the estimates of the value function. Recent works, such as \cite{vamvoudakis2021handbook, kerimkulov2020exponential, camilli2022rates, puterman1981convergence}, have explored iteration convergence rates in terms of the iteration number but do not consider the explicit effects of approximation error on performance.

This paper explores the effects of approximation error, and derives bounds on the error between the estimates of the value function and the corresponding control law. These bounds are explicit in terms of the number of bases $N$ used, and the geometric placement of centers that determines the bases.

\subsection{Summary of New Results} \label{sec:SummaryResults}
As is often carried out in RL \cite{lewis2013reinforcement,kamalapurkar2018reinforcement}, we can motivate the paper strategy by recalling the structure of PI. When the feedback function $\mu_i$ is known, we define the differential operator $A$  to be given by  $(A v)(x):= \left( f(x) + g(x) \mu(x) \right)^{\scriptscriptstyle\mathsf{T}} \nabla v(x)$ and $b(x) = -r(x,\mu(x))$. We then define $v_i$ as the solution to the partial differential equation 
\begin{align}%
(Av_i)(x)=b(x) &:=-r(x,\mu_i(x)), \label{eq:HamiltonianOperator} \\ 
v_i(0)&=0.\nonumber%
\end{align}%
When $v_i$ is determined from the above equation, we can subsequently define a new feedback law $\mu_{i+1}$ from the identity
\begin{equation} \label{eq:contrIter}
\mu_{i+1}(x)=-\frac{1}{2} R^{-1} g^\mathsf{T}(x) \nabla v_i(x).
\end{equation}
Setting $i\to i+1$ and repeating these steps generates a sequence of iterates 
 $\{(\mu_i,v_i)\}_{i\in \NN}$ that approximate the optimal functions  $\mu^*$ and $V^*$ that satisfy the HJB equations \cite{lewis2013reinforcement,abu2005nearly,bertsekas2012dynamic}.  

This paper derives {\it rates of convergence} for approximations of the solution $v$ of the partial differential equation $Av = b$, defined in \ref{eq:HamiltonianOperator}, for a given $\mu(x)$.  It also provides rates of convergence for the controller $\mu_{i+1}$ approximation error generated by the PI method. 
Under the hypothesis that the solution $v\in H$, where $H$ is a reproducing kernel Hilbert space (RKHS), we describe precise conditions on the reproducing kernel $\knl$ associated with $H$ that ensures 
\[
\|v-v_N\|_{H} \leq O\left ( \sup_{x\in \Omega} {\sqrt{\knl(x,x)-\knl_N(x,x)}}\right ).
\]
In the above inequality, $v_N$ is an approximate solution contained in the finite dimensional space $H_N:=\text{span}\{\knl(\cdot,\xi_i)\in H\ | \ \xi_i \in \Xi_N \}$ determined by the $N$ centers $\Xi_N:=\{\xi_1,\ldots,\xi_N\}\subset \Omega$. In this equation $\knl_N$ is the known reproducing kernel of $H_N$.  We emphasize the following:
\begin{enumerate}
    \item The above bound makes explicit the relationship of the center locations $\Xi_N$ to the error in solutions of the operator equation.
    \item For some popular kernels it is possible to bound the above expression in terms of the fill distance $h_{\Xi_N,\Omega}:=\sup_{x\in \Omega} \inf_{\xi_i\in \Xi_N}\|x-\xi_i\|$ of centers $\Xi_N$  in the set $\Omega$, 
    \[
    \|v-v_N\|_{H} \leq O(h^s_{\Xi_N,\Omega}),
    \]
    where $s$ is a parameter that measures the regularity of the kernel $\knl$. \label{sec:fillDistDef} 
    Thus, the rate of convergence of the approximation error depends on the {\it smoothness} of the basis and the {\it geometric distribution} of the centers in $\Xi_N\subset \Omega$ that define the basis. 

\end{enumerate}

\section{Theoretical Foundations}
\subsection{Symbols and Definitions}
In this paper $\RR$ and $\RR^+$ are the real numbers and nonnegative real numbers, respectively. The non-negative integers are denoted $\NN_0$, while the positive integers are $\NN$. When $U,V$ are normed vector spaces, $\calL(U,V)$ is the normed vector space of bounded linear operators from $U$ to $V$, and we just write $\calL(U)$ for $\calL(U,U)$. The range of an operator $T$ is denoted $R(T)$ and the nullspace of $T$ is written $N(T)$. The Lebesgue spaces $L^p(\Omega)$ are equipped with the usual norms 
\[
\|f\|_{L^p(\Omega)}:=
\left \{ \begin{array}{lll}
\left ( \int_\Omega |f(x)|^p dx  \right)^{1/p} &  1\leq p < \infty,& \\
 \text{ess sup}\{ |f(x)| \ | \ x \text{ a.e. in } \Omega\}  & p=\infty.&
\end{array}
\right . 
\]

\subsection{Reproducing Kernels and Native Spaces}
\label{sec:RKHSbackground}
A real-valued native space, denoted as $H(\Omega)$ over a set $\Omega$, is defined using a reproducing kernel $\knl(\cdot,\cdot):\Omega \times \Omega \rightarrow \RR$. This kernel, a Mercer kernel, is continuous, symmetric, and of positive type, which means that for any collection $\Xi_N \subset \Omega$ of $N$ points,  the Gramian matrix $\KK(\Xi_N,\Xi_N):=[\knl(\xi_i,\xi_j)]\in \RR^{N\times N}$ is positive semidefinite. Once such a kernel is selected, the native space $H(\Omega)$ is defined as the closed linear span of the kernel sections $\knl_x(\cdot):=\knl(x,\cdot)$, 
\begin{equation}
H(\Omega):=\overline{\text{span}\{\knl_x\ | \ x\in \Omega \}}. \label{eq:Hdef}
\end{equation}
A few properties of the evaluation functional $E_x:H(\Omega)\rightarrow \RR$ play a particularly important role in this paper. 
By definition, the evaluation functional satisfies $E_x f:=f(x)$ for all $f\in H(\Omega)$, and it is a bounded operator from $H(\Omega)\rightarrow \RR$. Every native space satisfies the  reproducing formula that connects the evaluation functional to inner products~via $E_xf=f(x)=(f,\knl_x)_H$ for all $f\in H(\Omega) , x\in \Omega. $
Moreover, since $E_x$ is a bounded operator, its adjoint $E_x^*:=(E_x)^*:\RR\rightarrow H(\Omega)$ is also a bounded linear operator. It is given by the formula 
$E_x^*\alpha:=\knl_x \alpha$ for all $\alpha \in \RR, x\in \Omega.$ 

In this paper, we always assume that the kernel $\knl(\cdot,\cdot)$ is bounded on the diagonal. That is, it is assumed that there is a $\bar{\knl}>0$ such that $\knl(x,x)\leq \bar{\knl}^2$ for all $x\in \Omega. $
This ensures that all the functions in $H(X)$ are bounded, and that the evaluation operator $E_x$ is uniformly bounded
$\|E_x\|\leq \bar{\knl}$ for all  $x \in \Omega$, and that we have the continuous embedding 
$H(\Omega)\hookrightarrow C(\Omega)$. Many popular kernels are bounded on the diagonal including the exponential, inverse multiquadric, Wendland, and Sobolev-Matérn kernels \cite{wendland}.

\subsubsection{Derivatives in Native Spaces}
\label{sec:nativederivs}
 
When $\knl$ is a Mercer kernel, having smoothness $\knl\in C^{2s}(\Omega \times \Omega)$ with $s \in \NN$, that defines the native space $H(\Omega)$, it is possible to express the action of the partial derivative operator $D^\alpha$ on functions in $H(\Omega)$  in terms of the partial derivatives of the kernel. Suppose we fix $y$ and are interested in partial derivatives with respect to $x$. To compute partial derivatives of the kernel, we  interpret a multiindex $\alpha=(\alpha_1,\ldots,\alpha_d,\alpha_{d+1},\ldots,\alpha_{2d})\in \NN_0^{2d}$  as having all zeros in the last $d$ entries, so that $\alpha:=(\alpha_1,\ldots,\alpha_d,0,\ldots, 0)\in \NN_0^{2d}$ and 
\begin{align*}
    &\left ( D^\alpha \knl\right)_x(y):=\left (D^\alpha \knl \right) (x,y)\\
    &:=\frac{\partial^{|\alpha|}}{\partial x_1^{\alpha_1},\cdots,\partial x_d^{\alpha_d}} \knl(x_1,\ldots,x_d,y_1,\ldots,y_d) \quad \forall x,y \in \Omega. 
\end{align*}
Theorem (1) from \cite{zhou2008derivative} specifies necessary conditions for the kernel, which entail it being a Mercer kernel that is sufficiently smooth, to prove that the derivative operator is a bounded operator on the native space and is vital to proving the results of this paper. Specifically, under the hypotheses described in the theorem, we have $\left ( D^\alpha h\right)(x)=\left ((D^\alpha \knl)(x,\cdot),h \right)_{H(\Omega)}=\left ( D^\alpha_x\knl,h\right)_{H(\Omega)}$ for all $x\in \Omega$, $h\in H(\Omega)$ and $\sum \alpha_i \leq s$.

%
%

\subsubsection{Approximation in Native Spaces}
\label{sec:approxRKHS}


To approximate function in $H(\Omega)$, we define 
$H_N:=\text{span}\{ \knl_{\xi_i}\ | \ \xi_i \in \Xi_N \}\subseteq H(\Omega) $
the space of approximants constructed using kernel sections defined in terms of the $N$ locations $\Xi_N\subset \Omega$. 
Let $\Pi_N$ be the $H(\Omega)$-orthogonal projection onto $H_N$. It is known that we have the general bound 
\[
\epsilon_{N,f}(x):= |E_x ( I-\Pi_N)f | \leq \Pwr_{H,N}(x) \|f\|_{H(\Omega)}  
\]
for all $f\in H(\Omega)$ and $x\in \Omega$, 
where the power function $\Pwr_{H,N}(x)$ is defined by 
$
\Pwr_{H,N}(x):=\sqrt{ \knl(x,x) - \knl_N(x,x)} \quad \text{ for all } x\in \Omega. 
$
The kernel $\knl_N(\cdot,\cdot)$ is the reproducing kernel of $H_N$ with
\begin{align}
\knl_N(x,y)&:=(\Pi_N \knl_x,\Pi_N \knl_y)_{H(\Omega)} \nonumber \\
&= \knl_{\Xi_N}^\mathsf{T}(x)\KK^{-1}(\Xi_N,\Xi_N)\knl_{\Xi_N}(y), \label{eq:knlN}
\end{align}
where $\knl_{\Xi_N}(\cdot)=\begin{bmatrix} \knl_{\xi_1}(\cdot) & \cdots & \knl_{\xi_n}(\cdot)\end{bmatrix}^\mathsf{T}$. 
This expression is used in a few different places in this paper.

%
%
\section{Offline Approximation in a Native Space}
\label{sec:criticNN}

\subsection{The Operator Framework in a Native Space} 
\label{sec:opframework}
We carry out value function approximation and subsequent analysis by first posing  \ref{eq:hamiltonian} as an operator equation. We define the differential operator $A$ as 
\begin{align*}
    (Av)(x):= \left ( f(x) + g(x) \mu(x)  \right )^\mathsf{T} \nabla v(x) \quad \text{ for all } x \in \Omega,
\end{align*}
whenever $v$ is sufficiently smooth. Note that the operator $\nabla$ in the above equation is defined in the usual way, with 
\begin{align*}
    \nabla f&:= \left ( \frac{\partial f}{\partial x_1}, \cdots  , \frac{\partial f}{\partial x_d}  \right)^\mathsf{T} :=\left ( D^{e_1} f,\cdots, D^{e_d}f\right)^\mathsf{T},
\end{align*}
where $D^{\alpha}(\cdot)$ is defined in Section \ref{sec:nativederivs} for any multiindex $\alpha\in \NN^d_0$ and $e_k$ is the canonical multiindex obtained by setting the $k^{th}$ entry to one and all other entries to zero. 

The next theorem expresses some mapping properties of the operator $A$ essential to our approximation schemes below. 
\begin{theorem}
\label{th:AandAstar}
Let the hypotheses of Theorem 1 in \cite{zhou2008derivative} hold  and further suppose that $\mu$ and $f_i,g_i$ for $1\leq i\leq d$ are multipliers for $H(\Omega)$. Then
\begin{enumerate}
    \item The  operator $A:H(\Omega)\rightarrow L^2(\Omega)$ is bounded, linear, and compact.
    \item The adjoint operator $A^*:L^2(\Omega)\rightarrow H(\Omega)$  has the representation \label{th:AandAstar2}
    \begin{align*}
        &\left ( A^* h \right)(y)
        := \int_\Omega \left ( \nabla_x \knl (y)\right)^\mathsf{T} \left (f(x) + g(x)\mu(x) \right) h(x)dx\\
        &:=\int_\Omega \ell^*(y,x)h(x)dx
    \end{align*}
    for any $y\in \Omega$ and $h\in L^2(\Omega)$. 
    \item Considered as an operator $A^*:L^2(\Omega)\rightarrow H(\Omega)$, the operator $A^*$ is compact. 
\end{enumerate}
\end{theorem}
\begin{proof}
\noindent (1.) It is clear that $A$ is linear. In part (1) of Theorem (1) from \cite{zhou2008derivative}, which is shown in the Appendix, we know that $\frac{\partial \knl(x,\cdot)}{\partial x_i} \in H(\Omega)$ for $x\in \Omega$ and $1\leq i\leq d$. From part (3) of the same Theorem, we have the continuous embedding $H(\Omega)\hookrightarrow C^s(\Omega)\hookrightarrow C(\Omega)$.
So for any $V\in H(\Omega)$,  we also know that $\frac{\partial V}{\partial x_k} \in C(\Omega)$ for $1\leq k\leq d$. This means that $x\mapsto (f(x)+g(x)\mu(x))^T\nabla V(x)$ is continuous since $f_i,g_i,\mu$ are multipliers for $C(\Omega)$.  We have 
\begin{align*}
    |(f(x)&+g(x)\mu(x),\nabla V(x))_{\RR^d}|^2 \\
    &\leq 
    \sum_{i=1}^d \|f_i+g_i\mu\|^2_{C(\Omega)} \sum_{i=1}^d\left \|\frac{\partial V}{\partial x_i}\right \|^2_{C(\Omega)} \\
    &\leq C \sum_{i=1}^d \|f_i+g_i\mu\|^2_{C(\Omega)} \|V\|^2_{H(\Omega)}.
\end{align*}
for a constant $C>0$ that comes from (3) of 
Theorem \ref{th:Zhou1} and the equivalence of norms on $\RR^d$. We conclude that $A:H(\Omega)\rightarrow C(\Omega)$ is a bounded linear operator.

\noindent(2.) 
The adjoint operator $A^*:L^2(\Omega)\rightarrow H(\Omega)$ of the bounded linear operator $A:H(\Omega)\rightarrow L^2(\Omega)$ is bounded and linear  by definition. Since $A$ is compact, $A^*:L^2(\Omega)\rightarrow H(\Omega)$ is compact  from Theorem 4.12 of \cite{kress1989linear}.  We only need to establish the representation. Using  (2) of Theorem \ref{th:Zhou1},  we have 
\begin{align*}
    (Av&,h)_{L^2(\Omega)}\\
    &=\int_\Omega \sum_{i=1}^d \left (f_i(x)+g_i(x)\mu(x) \right)(D^{e_i}_x\knl, v)_{H(\Omega)}h(x) dx, \\
    &=\left ( v, {\int_\Omega (\nabla_x\knl(x,\cdot))^T (f(x)+g(x)\mu(x))h(x)dx} \right )_{H(\Omega)},\\
    &=(v,A^*h)_{H(\Omega)}.
\end{align*}
We conclude that for all $y\in \Omega$ and $h\in L^2(\Omega)$, it holds that 
\begin{align*}
    (A^*h)(y):&=\int_\Omega (\nabla_x \knl)^T(y)(f(x)+g(x)\mu(x))h(x)dx\\
    &=\int_\Omega (\nabla_x \knl(x,y))^T(f(x)+g(x)\mu(x))h(x)dx
\end{align*}
where $(\nabla_x\knl)(y)=\{\partial \knl(x,y)/\partial x_1,\ldots,\partial\knl(x,y)/\partial {x_d} \}^T\in \RR^d$. 

We finally turn to the compactness of $A^*$ when we consider it as an operator from $L^2(\Omega)\rightarrow L^2(\Omega)$.  We  define the unsymmetric kernel function $\unl^*:\Omega\times \Omega \rightarrow \RR$ as 
\begin{align*}
\unl^*(y,x)&:=(A\knl_y)(x):=(\nabla_x\knl(x,y))^T(f(x)+g(x)\mu(x)), \\
&=\sum_{k=1}^d \frac{\partial \knl}{\partial x_i}(x,y)(f_i(x) + g_i(x) \mu(x)), \quad \forall x,y\in \Omega. 
\end{align*}
We also define its ``dual unsymmetric kernel'' as 
\begin{align*}
    \unl(x,y):=\unl^*(y,x) \quad \text{ for all } x,y\in \XX.
\end{align*}
When we define the unsymmetric section $\unl_y(\cdot)=\unl(\cdot,y)$, the   definition of $\ell$ is useful since 
\[
\ell_y(x):=\ell(x,y) = (A\knl_y)(x). 
\]
We also have the integral operator representation 
\begin{align}
(A^*h)(y):=\int_\Omega \unl^*(y,x)h(x)dx. \label{eq:elldef}
\end{align}
The boundedness of the map $A^*:L^2(\Omega)\rightarrow L^2(\Omega)$ follows immediately by continuity of the kernel $\unl$ since 
\[
\|A^*h\|_{L^2(\Omega)}^2 \leq \|\unl^*\|_{C(\Omega\times \Omega)}^2 \|h\|^2_{L^2(\Omega)}. 
\]
But by Theorem 2.27 of \cite{kress1989linear}, an integral operator from $L^2(\Omega)\rightarrow L^2(\Omega)$ with a continuous kernel  is compact.

\end{proof}

As discussed in section \ref{ssec_probstmt}, PI is based on the recursive solution of the operator equation  
$Av=b \in L^2(\Omega)$.
If $b\in R(A)$, the above equation has a solution, and if $N(A)={0}$, it is unique.  In any case   the operator $(A|_{N(A)^\perp})^{-1}:R(A)\to N(A)^{\perp}$ is well-defined. However, the operator  $(A|_{N(A)^\perp})^{-1}:N(A)^\perp \rightarrow (R(A),L^2(\Omega))$ is not bounded in general since $A:H(\Omega)\rightarrow L^2(\Omega)$ is compact. This complicates approximations.

A common way to approximate the solution of such an equation is to seek the minimum $v^*\in H(\Omega)$  of the offline optimization problem 
\begin{equation}
v^*=\text{argmin}_{v\in H(\Omega)} J(v):= \frac{1}{2} \|Av-b\|^2_{L^2(\Omega)}. \label{eq:leastsq1}
\end{equation}
When we rewrite the cost functional in the form
\begin{align*}
\frac{1}{2} &\|Av-b\|^2_{L^2(\Omega)} \\
&=\frac{1}{2} 
(A^*Av,v)_{H(\Omega)} - (v,A^*b)_{H(\Omega)} + \frac{1}{2}
(b,b)_{L^2(\Omega)}, 
\end{align*}
we can calculate its Frechet derivative $DJ(v):H(\Omega)\rightarrow H(\Omega)$ that satisfies 
\[
(DJ(v),w)_{H(\Omega)}:= (A^*Av-A^*b,w)_{H(\Omega)}=(A^*A\tilde{v},w)_{H(\Omega)}
\]
for all directions $w\in H(\Omega)$, with $A\tilde{v}:=Av-b$. Therefore, a minimizer satisfies the operator equation 
$A^*A v=A^*b$, or
\begin{align}
    \calA v&=y \label{eq:op2}
\end{align}
where  $\calA=A^*A:H(\Omega)\rightarrow H(\Omega)$ and $y=A^*b\in H(\Omega)$. 
 Offline approximations of the solution of the above operator equation  can be interpreted as approximations  of the pseudoinverse solution $V^*:=\calA^\dagger b \equiv (A^*A)^{-1}A^*b$. The pseudoinverse operator $\calA^\dagger$ is well-defined since $\calA$  is self-adjoint, compact, and nonnegative \cite{engl1996regularization}.

\subsection{Offline Approximations}
\label{sec:offapprox}
We now turn to the study of approximations of the solution of the operator equation \ref{eq:op2}. 
This operator equation is defined in terms of the bounded, linear, compact operator $\calA:H(\Omega)\rightarrow R(A^*):=W(\Omega)\subseteq H(\Omega) \subset L^2(\Omega)$. 
Since $y=A^*b\in R(A^*):=W(\Omega)$, \ref{eq:op2} always has a solution. It will be unique if $\calA$ is injective, and in this case $\calA^{-1}$ is a well-defined operator. However, when $\calA^{-1}$ exists it is generally not a bounded operator (unless $W(\Omega)$ is finite dimensional). 

Here we assume that bases used for approximation are defined in terms of kernel sections located at the $N$ centers $\Xi_N:=\{\xi_1,\ldots,\xi_N\}\subset \Omega$. We define the finite dimensional spaces of approximants 
\begin{align*}
    H_N&:=\text{span}\{ \knl_{\xi_i}(\cdot):=\knl(\cdot,\xi_i) \ | \ \xi_i \in \Xi_N\} \subset H(\Omega),\\
    L_N&:=\text{span}\{ \unl_{\xi_i}(\cdot):=\unl(\cdot,\xi_i) \ | \ \xi_i \in \Xi_N\}\subset L^2(\Omega), \\
    W_N&:=\text{span}\{ \wnl_{\xi_i}(\cdot) \ | \ \xi_i \in \Xi_N\} \subset W(\Omega):=R(A^*).
\end{align*}
We define $\unl(x,y):=\unl^*(y,x)$, where $l^*(y,x)$ is defined in Theorem \ref{th:AandAstar}. From this definition, we have $\ell_{\xi_i}(x):=(A\knl_{\xi_i})(x)$. So these bases satisfy the relations 
$\unl_{\xi_i}=A\knl_{\xi_i}$, and $
    \wnl_{\xi_i}=A^*\unl_{\xi_i}=A^*A\knl_{\xi_i}$.
We denote by $\Pi_N:H(\Omega)\rightarrow H_N$ the projection of $H(\Omega)$ onto $H_N$. 
We define the Galerkin approximation $v_N\in H_N$ of the solution $v\in H(\Omega)$ of \ref{eq:op2} to be given by $v_N:=(\Pi_N \calA|_{H_N})^{-1}\Pi_N y:=G_Ny$. This is equivalent to the variational equations 
\begin{align}
\left (\calA v_N - y,\knl_{\xi_i} \right )_{H(\Omega)} &=0 \quad \text{ or, } \label{eq:gal1} \nonumber \\
\left (A^*A v_N - A^*b,\knl_{\xi_i} \right )_{H(\Omega)} &=0 
\quad \text{ for } 1\leq i\leq N.  \nonumber
\end{align}
It is also worth noting that the Galerkin solution $v_N$ above coincides with the Galerkin approximation of $Av=b$ in 
\[
\left (Av_N-b,\ell_{\xi_i} \right)_{L^2(\Omega)}=0 \quad \text{ for } 1\leq i \leq N.
\]

\subsection{Coordinate Realizations}
\label{sec:coordreal}
The study of the rates of convergence of the above approximations utilize coordinate representations of the operators. 
We need representations of the operator $A^*A:H(\Omega)\rightarrow H(\Omega)$.  For $A^*A$ we have 
\begin{align*}
    (A&^*Av)(y):=\int_\Omega \unl^*(y,x)(\unl^*(\cdot,x),v)_{H(\Omega)} dx. \\
    &= \int_\Omega \left ( \nabla_x\knl(x,y)^\mathsf{T} \psi(x) \psi(x)^\mathsf{T} \nabla_x\knl(x,\cdot),v\right )_H dx, 
\end{align*}
where $\psi(x):=f(x)+g(x)\mu(x)$. 

The representation of the operators $A^*A$ can now be used to determine the coordinate representations of the Galerkin approximations above.
Define the matrix 
\begin{align*}
\Phi(x,\Xi_N)&:=\begin{bmatrix}
\frac{\partial \knl(x,\xi_1)}{\partial x_1} & \cdots & \frac{\partial \knl(x,\xi_N)}{\partial x_1} \\
\vdots &   & \vdots \\
\frac{\partial \knl(x,\xi_1)}{\partial x_d} & \cdots & \frac{\partial \knl{(x,\xi_N)}}{\partial x_d}
\end{bmatrix}\in \RR^{d\times N}.
\end{align*}
Then for any two functions $v_N,w_N\in H_N$ with $v_N:=\sum_{j=1}^N \alpha_j \knl_{\xi_j}$ and $w_N:=\sum_{k=1}^N \beta_k \knl_{\xi_k}$ , we have 
\begin{align}
&\left ( A^*A v_N,w_N\right )_{H(\Omega)} \label{eq:coord1}\\ \nonumber
&\hspace*{.2in} = \beta^\mathsf{T} \underbrace{\left ( 
\int_\Omega \Phi(x,\Xi_N)^\mathsf{T}\psi(x)\psi(x)^\mathsf{T}\Phi(x,\Xi_N) dx \right)}_{[\int \unl^*(x,\xi_i)\unl(x,\xi_j)dx]}
\alpha 
\end{align}
with $\alpha:=[\alpha_1,\ldots,\alpha_N]^\mathsf{T}\in \RR^N$, $\beta:= [\beta_1,\ldots,\beta_N]^\mathsf{T}\in \RR^N$.
\subsection{Offline Rates of Convergence}

\begin{theorem}
\label{th:offlinerate}
    Let the hypothesis of Theorem \ref{th:AandAstar} hold, and suppose that the unknown value function $v$ satisfies the regularity condition $v=\mathcal{K} q$ for some fixed $q\in L^2(\Omega)$ where $\mathcal{K}:L^2(\Omega) \rightarrow H$  is the integral operator 
    $
    v(x)=\int_{\Omega}\knl(x,\eta)q(\eta)d\eta, 
    $
 and that the choice of centers $\Xi_N$ ensures that an ideal ``offline'' persistence of excitation (PE) condition holds for the offline Galerkin approximations above. That is,  there is a constant $\beta(N)>0$ such that 
    \[
\beta(N) I_N \leq \int_\Omega \Phi(x,\Xi_N)^\mathsf{T} \psi(x) \psi(x)^\mathsf{T} \Phi(x,\Xi_N) dx 
    \] 
where $I_N$ is the identity matrix on $\RR^N$. 
    Then the solution $v_N$ of the Galerkin equations   exists and is unique for all $N\in \NN$. If the Galerkin method is convergent, then  there is a constant $C>0$ such that the solution $v_N$ satisfies 
    the error estimate 
    \begin{align*}
        \|v-v_N\|_{H(\Omega)} &\leq C \sup_{\xi\in \Omega}\Pwr_{H,N}(\xi) \|v\|_{H(\Omega)} \\
        &= C \sup_{\xi\in \Omega}\sqrt{\knl(\xi,\xi)-\knl_N(\xi,\xi)}\|\mathcal{K}^{-1}v\|_{L^2(\Omega)}.
    \end{align*}
\end{theorem}
\begin{proof}
When we write $v_N:=\sum_{j=1}^N\alpha_i \knl_{\xi_i}$, the Galerkin approximations give rise to the matrix equations

\begin{align}
\left [\int_\Omega \Phi(x,\Xi_N)^\mathsf{T}\psi(x)\psi(x)^\mathsf{T} \Phi(x,\Xi_N) dx \right ] \alpha &= \begin{Bmatrix} (A^*b)(\xi_1) \\ \vdots \\ (A^*b)(\xi_N)\end{Bmatrix} \notag  \\
=\int_\Omega \Phi(x,\Xi_N)^\mathsf{T} \psi(x)\ b\ &dx \label{eq:matrixEquations}
\end{align} 
with $\alpha=[\alpha_1,\ldots,\alpha_N]^\mathsf{T} \in \RR^N$. 
    The representation in \ref{eq:coord1} makes clear that the offline PE condition ensures that the coefficient matrix is invertible. Also, the  operator $G_N \calA:=(\Pi_N \calA|_{H_N})^{-1} \Pi_N \calA$ is a projection onto $H_N$ since for any $p_N\in H_N$, we have 
    \[
G_N \calA p_N=(\Pi_N \calA|_{H_N})^{-1} \Pi_N \calA|_{H_N} p_N=p_N.
     \]
    From the triangle inequality we have the pointwise bound 
    \begin{align*}
        \|v-v_N\|_{H} & \leq \|v- G_N\calA p_N\|_H + \|G_N\calA p_N - G_N\calA v\|_H \\ 
        &\leq \|v-p_N)\|_H + \|G_N\calA (v-p_N)\|_H \\
        &\leq (1+\tilde{C}) \|v-p_N\|_H
    \end{align*}
    for any $p_N\in H_N$. In this inequality we have used the fact that in  a convergent Galerkin scheme the matrix $G_N\calA$ is uniformly bounded in $N$: there is a constant $\tilde{C}>0$ such that  $\|G_N\calA\|\leq \tilde{C}$ for all $N>0$ \cite{kress1989linear}. We choose $p_N:=\Pi_N v$. The theorem now follows from the characterizations of projection/interpolation errors in terms of the power function in a native space discussed in Section \ref{sec:approxRKHS}. We have 
    \begin{align*}
        \|v-v_N\|_{H} & \leq  (1+\tilde{C}) \|(I-\Pi_N)v\|_H \\
        &\leq (1+\tilde{C}) \|\Pwr_{H,N}\|_{L^2(\Omega)} \|q\|_{L^2(\Omega)}. 
    \end{align*}
    The last line stems from the proof of Theorem 11.23 in Section 11.5 of \cite{wendland}. Alternatively, we have 
    \begin{align*}
    \|v-v_N\|_H&\leq (1+\tilde{C}) \sqrt{|\Omega|} \sup_{\xi\in \Omega} \Pwr_{H,N}(\xi) \|q\|_{L^2(\Omega)}
    \end{align*}
    for all $v=\mathcal{K}q$ with $q\in L^2(\Omega)$. 
\end{proof}

\medskip

\noindent \underline{Observations}:
We make several observations about how the result above compares to existing results. 

\medskip 

\noindent (1) We say that the offline PE condition in Theorem \ref{th:offlinerate} is ideal since it involves the integration over $\Omega$ that cannot usually be carried out in closed form. 

\medskip

\noindent (2) It is important to allow that the constant $\beta(N)$ in the offline PE condition above depends on the dimension $N$. This form of the PE condition could alternatively be written as 
\begin{align*}
    \hat{\beta}(N) \|v\|_{H(\Omega)}^2 \leq (A^*A v,v)_{H(\Omega)} \quad \text{ for all } v\in H_N 
\end{align*}
for another constant $\hat{\beta}(N)$ that depends on $N$. But we know that $A^*A:H(\Omega)\rightarrow H(\Omega)$ is compact from Theorem \ref{th:AandAstar} above. If the PE condition above holds for a constant $\hat{\beta}$ that does not depend on $N$, we could conclude that $(A^*A)^{-1}$ is a bounded linear operator. But since $A^*A$ is compact, this is only true when $H(\Omega)$ is finite dimensional. In general, we must allow that the lower bound in the ideal  PE condition depends on $N$. 

\medskip

\noindent (3) The right hand side in the above error bound is explicit since we know $\knl_N$ as given in \ref{eq:knlN}.

\medskip 

\noindent (4) Using normalized regressors is popular practice, as summarized in \cite{kamalapurkar2018reinforcement,lewis2013reinforcement}. This is useful when regressors may be unbounded, such as when using polynomial regressors \cite{abu2005nearly,vamvoudakis2010online}.
For the sake of obtaining simple analysis and error bounds, we do not use the normalized form. Here, regressors are always bounded when the RKHS $H(\Omega)$ is defined in terms of a kernel $\knl(\cdot,\cdot)$ that is bounded on the diagonal. We also assume that the controller $\mu$ that is implicit in the operator equation $Av=b$ generates a trajectory that lies in the compact set $\Omega$. Again, this choice is made for illustrating strong error bounds in the simplest possible form. 

For some standard kernel spaces, the error bounds in Theorem \ref{th:offlinerate} can alternatively be bounded from above in terms of the fill distance $h_{\Xi_N,\Omega}$  of centers $\Xi_N$ in $\Omega$, which is defined in section  \ref{sec:SummaryResults}.  
\begin{corollary}
\label{cor:cor1}
    Let the hypothesis in Theorem \ref{th:offlinerate} hold and further suppose that the kernel $\knl$ that defines $H$ is given as in Table 11.1 of \cite{wendland} or Table 1 of \cite{schaback94}.  Then if the domain $\Omega$ is sufficiently smooth, we have 
    \[
    \|v-v_N\|_{H} \leq O\left ( \sqrt{\calF(h_{\Xi_N})}\right )
    \]
    for a known function $\calF$ defined in Table 11.1 of \cite{wendland} or Table 1 of \cite{schaback94}. 
\end{corollary}
\begin{proof}

    From Theorem \ref{th:offlinerate}, we have that
    \begin{align*}
        \|v-v_N\|_{H} &\leq (1+\tilde{C}) \|\Pwr_{H,N}\|_{L^2(\Omega)} \|q\|_{L^2(\Omega)}\\
        &= \hat{C} \|\Pwr_{H,N}\|_{L^2(\Omega)} \|q\|_{L^2(\Omega)}. 
    \end{align*}

    But from Table 11.1 in \cite{wendland}, we have that
    \begin{align*}
        \Pwr_{H,N} &\leq \hat{C} \calF(h_{\Xi_N}))\\
        \|\Pwr_{H,N}\|&\leq |\hat{C}|\sqrt{\|\calF(h_{\Xi_N}))\|_{L^2(\Omega)}}\\
        \text{which implies,}\\
         \|v-v_N\|_{H} &\leq O\left(\sqrt{\calF(h_{\Xi_N})}\right)%
    \end{align*}%
\end{proof}%
For instance, for the Sobolev-Matern kernels of smoothness $r>0$, as  used in the numerical examples, we have 
\begin{equation}
\|v-v_N\|_{L^\infty(\Omega)} \leq    \|v-v_N\|_{H} \leq O\left ( {h^{\nu-d/2}_{\Xi_N}}\right ), \label{eq:maternBound}
\end{equation}
where $\nu$ is a smoothness parameter and $d$ is the dimension of the space in which $\Omega$ is contained. Thus, the approximation error converges at a rate that is bounded above by the fill distance raised to the smoothness parameter. The following theorem links the value approximation error with the controller approximation error

\begin{theorem} 
\label{th:controlBounds}
    Under the same assumptions in Theorem \ref{th:offlinerate} and Corollary \ref{cor:cor1}, for the next estimate $\mu_{i+1,N}$ of the control policy iteration $\mu_{i+1}$ in equation \ref{eq:contrIter}, we have 
    \begin{equation} \label{eq:controlBounds}
        \|   \mu_{i+1,N} - \mu_{i+1} \|_{C(\Omega)} \leq \gamma \|   v_{i,N} - v_i \|_H \leq O\left ( \sqrt{\calF(h_{\Xi_N})}\right ),
    \end{equation}
    where $\gamma$ is a constant that depends on the kernel choice and the set of centers. \label{th:controlBound}
\end{theorem}
\begin{proof}
$$
    \begin{aligned}
        & \text{Let } \tilde v = v-v_N \\
        & \| \mu_{i+1,N} - \mu_{i+1}  \|_{C(\Omega)} = \left\| \frac{1}{2} R^{-1} g^{\top} (\nabla v_N - \nabla v)\right\|_{C(\Omega)}\\
        & =\left\| \frac{1}{2} R^{-1} g^{\top} \nabla \tilde{v}\right\|_{C(\Omega)}\\
        & = \|\sum_{j, k} R_{i j}^{-1} g_{j k}{(\cdot)}\nabla \tilde{v}(\cdot)\|_{C(\Omega)}\\
        & \leq \sum_{j, k}  |R_{i j}^{-1}|\  \|g_{j k}{(\cdot)}\|_{C(\Omega)} \ \|\nabla \tilde{v}(\cdot)\|_{C(\Omega)}\\
        &\text{Theorem 1 from \cite{zhou2008derivative}, gives us:}\\
        & \| v \|_{C^1(\Omega)} =  \| v \|_{C(\Omega)} + \| \nabla v \|_{C(\Omega)} \leq \\
        & \| v \|_{C(\Omega)} + \max_k \| \frac{\partial v}{\partial x_k} \|_{C(\Omega)} \leq c \|  v \|_{H} \text{ for some constant c}\\
        &\text{Which implies} \\
        &\| \mu_{i+1,N} - \mu_{i+1} \|_{C(\Omega)} \leq\\ 
        &\gamma \sum_{j, k}  |R_{i j}^{-1}|\  \|g_{j k}{(\cdot)}\|_{C(\Omega)} \ \| \tilde{v}(\cdot)\|_{C(\Omega)}
        \leq  O\left ( \sqrt{\calF(h_{\Xi_N})}\right ) 
     \end{aligned} 
    $$
\end{proof}%

\section{Numerical Simulations}

In this section, we consider the nonlinear system shown in \cite{vamvoudakis2010online}:
\[\dot{x}=f(x)+g(x) u, \quad x \in R^2\]%
where
$$
\begin{aligned}
& f(x)=\left[\begin{array}{c}
-x_1+x_2 \\
-0.5 x_1-0.5 x_2\left(1-\left(\cos \left(2 x_1\right)+2\right)^2\right) 
\end{array}\right] \\
& g(x)=\left[\begin{array}{c}
0 \\
\cos \left(2 x_1\right)+2
\end{array}\right] 
\end{aligned}
$$
Using the typical cost function J associated with the linear quadratic regulator problem, we choose $R=1$ and $Q=I_2$, that is, the $2\times2$ identity matrix. With this cost function, the value function is
$V^*(x) = 0.5 x_1^2 + x_2^2$, and the optimal control policy is $u^*(x) = -(\cos(2x_1) + 2)x_2$.  The simulations presented in \cite{vamvoudakis2010online} use polynomial bases whose finite dimensional span contains the unknown value function. Here, we use the RKHS bases to illustrate the theoretical results of this paper. Certainly, the theoretical bounds extend to cases where the value function is not spanned by a finite number of polynomial bases functions. Using \ref{eq:matrixEquations} and a quadrature approximation, we solve for the coefficients $\alpha$. 

\begin{figure}[!tb]
    \centering
    \includegraphics[width = 0.43\textwidth]{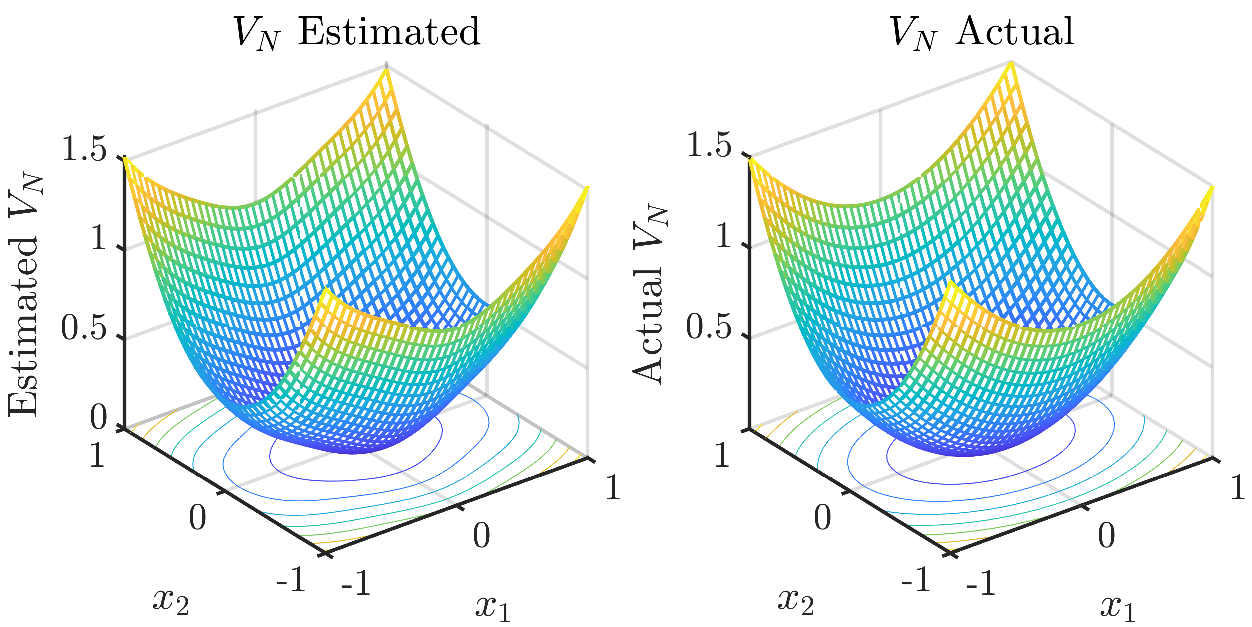}
    \caption{The estimated and ideal value functions over the spatial domain.}
    \label{fig:valueFuncPlot}
\end{figure}

\begin{figure}[!tb]
    \centering
    \includegraphics[width = 0.42\textwidth]{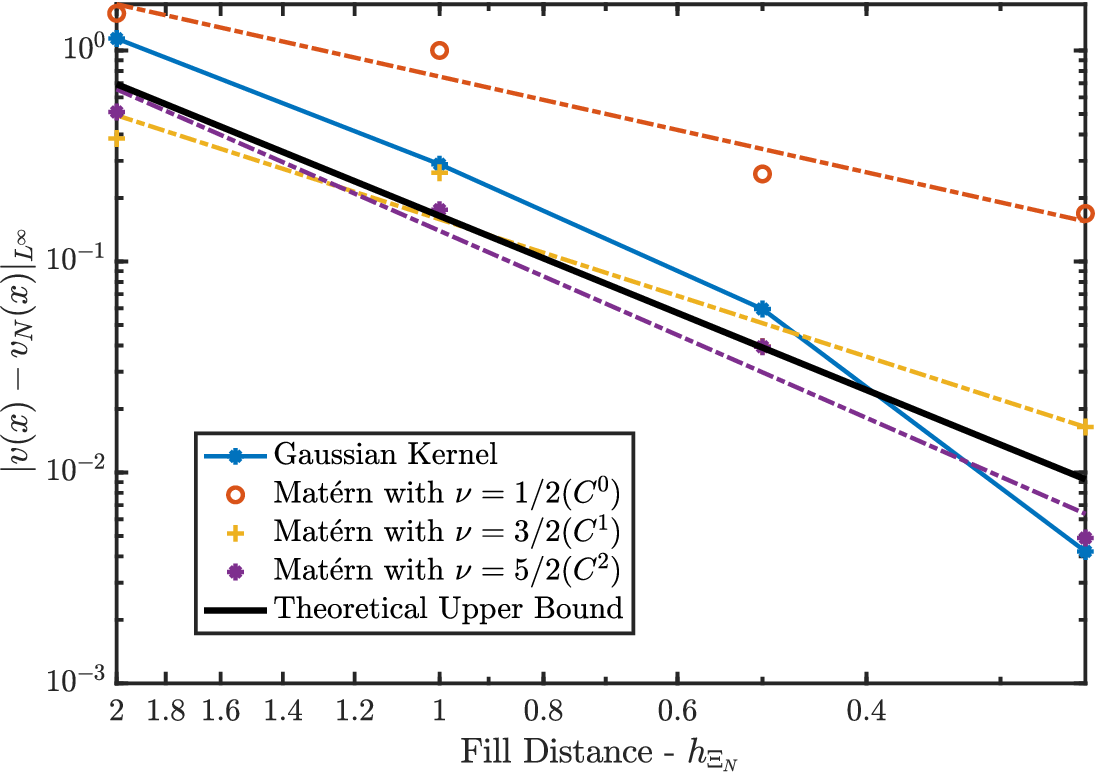} \caption{Plot depicting the value function approximation error decay for Gaussian and Matérn kernels. The linear segments in the plot correspond to fitting a straight line to the logarithm of the data.} \label{fig:decayPlot}%
\end{figure}

Then, the value function is approximated using $v_N:=\sum_{j=1}^N\alpha_i \knl_{\xi_i}$, with Gaussian and Matérn kernels as defined in \cite{williams2006gaussian}. The simulations utilized routines provided in \cite{Schaback2011MATLAB} for kernel based computations. The ideal control law is assumed to be known and is employed in this approximation. Our primary focus is to assess the accuracy of the value function approximation in an offline manner and to validate expected convergence rates. As shown in Fig. \ref{fig:valueFuncPlot}, the approximated value function closely matches the optimal one. In Fig. \ref{fig:decayPlot}, we see that as we decrease the fill distance (increase the number of centers), the approximation error decays as expected. 
Recall that the presented theoretical results apply to kernels of $C^{2s}$ smoothness, which means it applies to the case with $\nu = 5/2$ only (refer to \cite{williams2006gaussian}). This is validated by the fact that the line corresponding to $\nu = 5/2$ in the figure is steeper than the theoretical upper bound described in \ref{eq:maternBound}.

\begin{figure}[!tb]
    \centering
    \includegraphics[width = 0.41\textwidth]{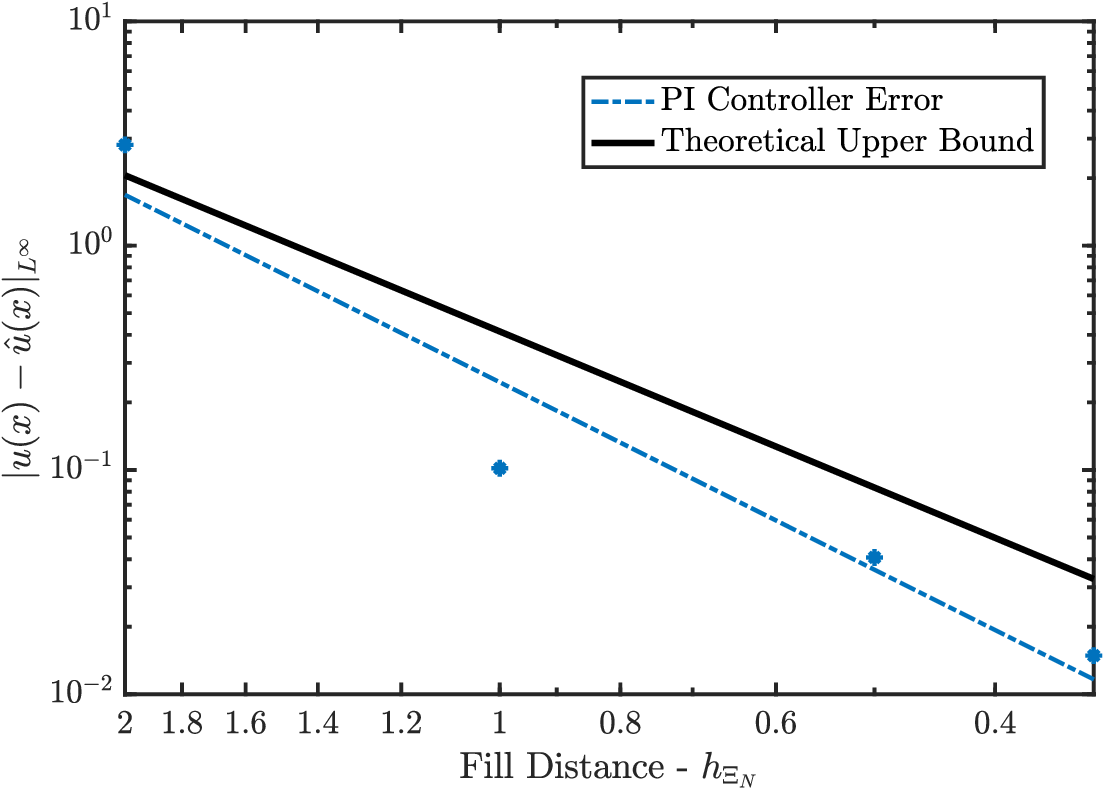}
    \caption{In this plot, the PI algorithm was used to calculate the error between the estimated $\hat u(x)$ and optimal $u(x)$ for different fill distances.  The rate of error decay is bounded from above by the theoretical limit.}
    \label{fig:controlDecay}
\end{figure}

\begin{figure}[!tb]
    \centering
    \includegraphics[width = 0.43\textwidth]{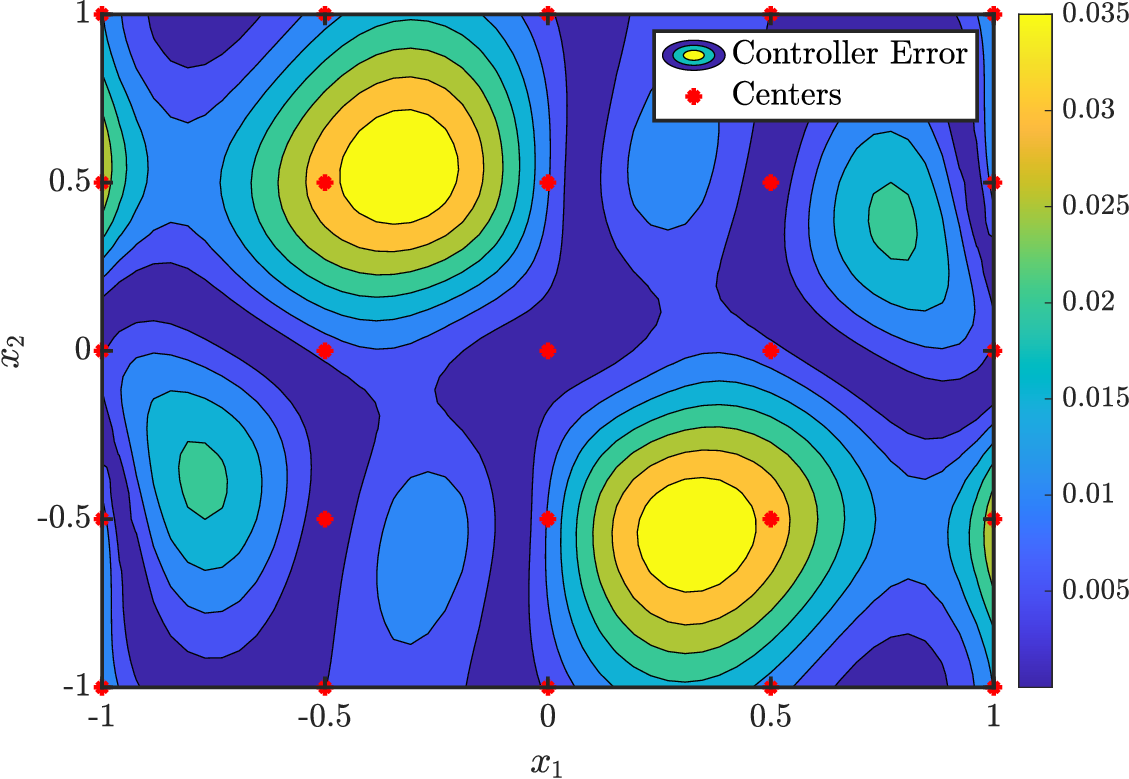}
    \caption{A geometric representation of the error with centers. Errors are consistently smallest at the centers and are largest away from the centers.}
    \label{fig:ControllerErr}
\end{figure}

\begin{figure}[!tb]
    \centering
    \includegraphics[width = 0.43\textwidth]{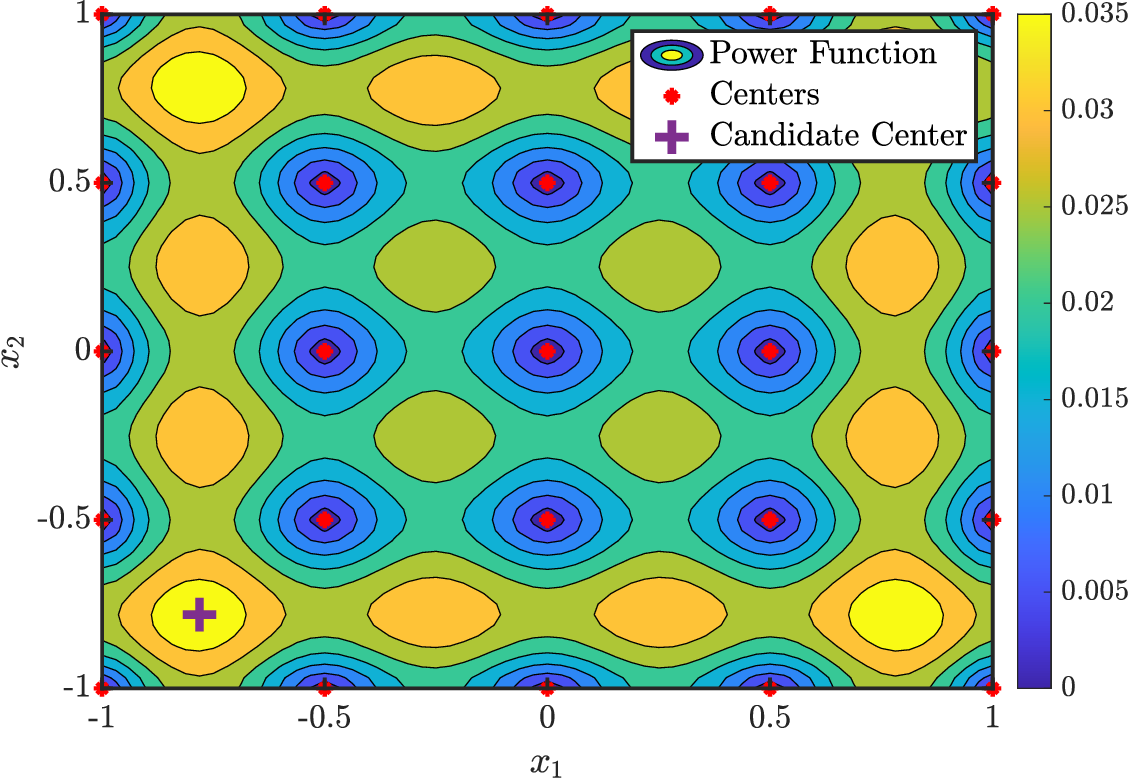}
    \caption{Power function plot with the centers, and a depiction of where a candidate center, to be augmented, may be chosen based on the maximum of the power function over the domain.}
    \label{fig:powerFuncPlot}
\end{figure}

Now, we begin with a stabilizing controller $\mu(x)$ and apply PI to approximate the optimal controller. Matérn kernel with $\nu=5/2$ is used in these simulations. Furthermore,  Fig. \ref{fig:controlDecay} shows the error between the ideal controller and the estimated controller is displayed for different fill distances. Again, the rate of error decay respects the limit predicted by \ref{eq:controlBounds}. Fig. \ref{fig:ControllerErr} is a geometric representation of the controller error plotted alongside the distribution of the centers. It is noteworthy that the error is generally smallest at the centers, and largest away from them.  Based on the results of Theorem \ref{th:controlBound}, one method to increase the number of bases adaptively is to position the next center at a location where the power function is largest. In Fig. \ref{fig:powerFuncPlot}, the power function and a candidate new basis are plotted with the centers. %


\section{Conclusion}
    In conclusion, this paper studies convergence rates for  value function approximations that arise in a collection of RKHS. These rates can help in practical scenarios such as determining the number and placement of basis functions to achieve the required accuracy. 
    These rates can also serve as the foundation for studies on rates of convergence for online actor-critic and RL methods.  
    Future directions include developing bases adaption techniques based on the error estimates presented in this work.

\section{Appendix}
\label{sec:Appendix}

The following theorem from \cite{zhou2008derivative} is key to the developments in this paper. 

\begin{theorem}[Zhou \cite{zhou2008derivative}, Theorem 1]
\label{th:Zhou1}
Let $\Omega \subset \RR^d$ be a connected  compact set that is equal to the closure of its nonempty interior, and let $\knl:\Omega \times \Omega \rightarrow \RR$ be Mercer kernel having smoothness $\knl \in C^{2s}(\Omega\times \Omega)$ for $s\geq 1$ that defines  the native space $H(\Omega)$. Then we have the following:
\begin{enumerate}
    \item For any $x\in \Omega$ and multiindex $|\alpha|\leq s$, it holds that $(D^\alpha \knl)_x(\cdot):=D_x^{\alpha}\knl(x,\cdot) =(D^\alpha\knl)(x,\cdot)\in H(\Omega)$.
    \item We have a pointwise representation of partial derivatives: for all $x\in \Omega$ and $h\in H(\Omega)$ we have 
    \begin{align*}
        \left ( D^\alpha h\right)(x)=\left ((D^\alpha \knl)(x,\cdot),h \right)_{H(\Omega)}=\left ( D^\alpha_x\knl,h\right)_{H(\Omega)}.
    \end{align*}
    \item We have the continuous embedding $H(\Omega)\hookrightarrow C^s(\Omega)$, with the norm bound 
    \begin{align*}
    \|h\|_{C^s(\Omega)} \leq \sqrt{d^m \|\knl\|_{C^{2s}(\Omega\times \Omega}}\|h\|_{H(\Omega)}.
    \end{align*}
\end{enumerate}
\end{theorem}

\bibliographystyle{IEEEtran}
\bibliography{bib}

\begin{thebibliography}{10}
\providecommand{\url}[1]{#1}
\csname url@samestyle\endcsname
\providecommand{\newblock}{\relax}
\providecommand{\bibinfo}[2]{#2}
\providecommand{\BIBentrySTDinterwordspacing}{\spaceskip=0pt\relax}
\providecommand{\BIBentryALTinterwordstretchfactor}{4}
\providecommand{\BIBentryALTinterwordspacing}{\spaceskip=\fontdimen2\font plus
\BIBentryALTinterwordstretchfactor\fontdimen3\font minus
  \fontdimen4\font\relax}
\providecommand{\BIBforeignlanguage}[2]{{%
\expandafter\ifx\csname l@#1\endcsname\relax
\typeout{** WARNING: IEEEtran.bst: No hyphenation pattern has been}%
\typeout{** loaded for the language `#1'. Using the pattern for}%
\typeout{** the default language instead.}%
\else
\language=\csname l@#1\endcsname
\fi
#2}}
\providecommand{\BIBdecl}{\relax}
\BIBdecl

\bibitem{bertsekas2012dynamic}
D.~Bertsekas, \emph{Dynamic Programming and Optimal Control: {{Volume
  I}}}.\hskip 1em plus 0.5em minus 0.4em\relax {Athena scientific}, 2012,
  vol.~1.

\bibitem{bertsekas1997nonlinear}
D.~P. Bertsekas, ``Nonlinear programming,'' \emph{Journal of the Operational
  Research Society}, vol.~48, no.~3, pp. 334--334, 1997.

\bibitem{lewis2013reinforcement}
F.~L. Lewis and D.~Liu, \emph{Reinforcement Learning and Approximate Dynamic
  Programming for Feedback Control}.\hskip 1em plus 0.5em minus 0.4em\relax
  {John Wiley \& Sons}, 2013.

\bibitem{bea1998successive}
R.~W. Bea, ``Successive {{Galerkin}} approximation algorithms for nonlinear
  optimal and robust control,'' \emph{International Journal of Control},
  vol.~71, no.~5, pp. 717--743, 1998.

\bibitem{beard1997galerkin}
R.~W. Beard, G.~N. Saridis, and J.~T. Wen, ``Galerkin approximations of the
  generalized {{Hamilton-Jacobi-Bellman}} equation,'' \emph{Automatica},
  vol.~33, no.~12, pp. 2159--2177, 1997.

\bibitem{abu2005nearly}
M.~{Abu-Khalaf} and F.~L. Lewis, ``Nearly optimal control laws for nonlinear
  systems with saturating actuators using a neural network {{HJB}} approach,''
  \emph{Automatica}, vol.~41, no.~5, pp. 779--791, 2005.

\bibitem{vamvoudakis2010online}
K.~G. Vamvoudakis and F.~L. Lewis, ``Online actor\textendash critic algorithm
  to solve the continuous-time infinite horizon optimal control problem,''
  \emph{Automatica}, vol.~46, no.~5, pp. 878--888, 2010.

\bibitem{vamvoudakis2021handbook}
K.~G. Vamvoudakis, Y.~Wan, F.~L. Lewis, and D.~Cansever, \emph{Handbook of
  Reinforcement Learning and Control}.\hskip 1em plus 0.5em minus 0.4em\relax
  {Springer}, 2021.

\bibitem{kiumarsi2017optimal}
B.~Kiumarsi, K.~G. Vamvoudakis, H.~Modares, and F.~L. Lewis, ``Optimal and
  autonomous control using reinforcement learning: {{A}} survey,'' \emph{IEEE
  transactions on neural networks and learning systems}, vol.~29, no.~6, pp.
  2042--2062, 2017.

\bibitem{kamalapurkar2018reinforcement}
R.~Kamalapurkar, P.~Walters, J.~Rosenfeld, and W.~Dixon, \emph{Reinforcement
  Learning for Optimal Feedback Control}.\hskip 1em plus 0.5em minus
  0.4em\relax {Springer}, 2018.

\bibitem{kerimkulov2020exponential}
B.~Kerimkulov, D.~Siska, and L.~Szpruch, ``Exponential convergence and
  stability of howard's policy improvement algorithm for controlled
  diffusions,'' \emph{SIAM Journal on Control and Optimization}, vol.~58,
  no.~3, pp. 1314--1340, 2020.

\bibitem{camilli2022rates}
F.~Camilli and Q.~Tang, ``Rates of convergence for the policy iteration method
  for mean field games systems,'' \emph{Journal of Mathematical Analysis and
  Applications}, vol. 512, no.~1, p. 126138, 2022.

\bibitem{puterman1981convergence}
M.~Puterman, ``On the convergence of policy iteration for controlled
  diffusions,'' \emph{Journal of Optimization Theory and Applications},
  vol.~33, pp. 137--144, 1981.

\bibitem{wendland}
H.~Wendland, \emph{Scattered Data Approximation}.\hskip 1em plus 0.5em minus
  0.4em\relax {Cambridge university press}, 2004, vol.~17.

\bibitem{zhou2008derivative}
D.-X. Zhou, ``Derivative reproducing properties for kernel methods in learning
  theory,'' \emph{Journal of computational and Applied Mathematics}, vol. 220,
  no. 1-2, pp. 456--463, 2008.

\bibitem{kress1989linear}
R.~Kress, V.~Maz'ya, and V.~Kozlov, \emph{Linear Integral Equations}.\hskip 1em
  plus 0.5em minus 0.4em\relax {Springer}, 1989, vol.~82.

\bibitem{engl1996regularization}
H.~W. Engl, M.~Hanke, and A.~Neubauer, \emph{Regularization of Inverse
  Problems}.\hskip 1em plus 0.5em minus 0.4em\relax {Springer Science \&
  Business Media}, 1996, vol. 375.

\bibitem{schaback94}
R.~Schaback, ``Error estimates and condition {{Numbers}} for radial basis
  function interpolation,'' \emph{Advances in Computational Mathematics},
  vol.~3, pp. 251--264, 1994.

\bibitem{williams2006gaussian}
C.~K. Williams and C.~E. Rasmussen, \emph{Gaussian Processes for Machine
  Learning}.\hskip 1em plus 0.5em minus 0.4em\relax {MIT press Cambridge, MA},
  2006, vol.~2.

\bibitem{Schaback2011MATLAB}
R.~Schaback, ``{{MATLAB}} programming for kernel based methods,'' unpublished,
  2011, technical report.

\end{thebibliography}
\end{document}